\documentclass[a4paper,10pt,twocolumn]{article}
\usepackage[utf8]{inputenc}
\usepackage[english]{babel}
\usepackage{amsmath}
\usepackage{units}
\usepackage{graphicx}
\usepackage{color}
\usepackage{authblk}
\usepackage{microtype}
\usepackage{abstract}
\AtBeginDocument{}
\usepackage[nomove]{cite}

\title{Directional Emission from Dielectric Leaky-Wave Nanoantennas}

\title{Directional Emission from Dielectric Leaky-Wave Nanoantennas}
\author[1]{Manuel Peter}
\author[2]{Andre Hildebrandt}
\author[3]{Christian Schlickriede}
\author[1]{Kimia Gharib}
\author[3]{Thomas Zentgraf}
\author[2]{Jens Förstner}
\author[1,*]{Stefan Linden}
\affil[1]{University of Bonn, Physikalisches Institut, D-53115, Bonn, Germany}
\affil[2]{University of Paderborn, Department of Electrical Engineering, D-33098, Paderborn, Germany}
\affil[3]{University of Paderborn, Department of Physics, D-33098, Paderborn, Germany}
\affil[*]{linden@physik.uni-bonn.de}

\begin{document}
	\twocolumn[ 
	\begin{center}
		{\Large \bf Directional Emission from Dielectric Leaky-Wave Nanoantennas}
		
	\end{center}
	\begin{center}
		{\large M. Peter,\textsuperscript{1} A. Hildebrandt,\textsuperscript{2} C. Schlickriede,\textsuperscript{3} K. Gharib,\textsuperscript{1} T. Zentgraf,\textsuperscript{3} J. Förstner,\textsuperscript{2} and S. Linden\textsuperscript{1,*}}
		
		\vspace{0.1cm}
		
		\textsuperscript{1}\textit{University of Bonn, Physikalisches Institut, D-53115, Bonn, Germany}\\
		\textsuperscript{2}\textit{University of Paderborn, Department of Electrical Engineering, D-33098
			Paderborn, Germany}\\
		\textsuperscript{3}\textit{University of Paderborn, Department of Physics, D-33098, Paderborn, Germany}\\
		\textsuperscript{*}\textit{linden@physik.uni-bonn.de}
	\end{center}
	\vspace{-1.3cm}
	\begin{onecolabstract}	
		An important source of innovation in nanophotonics is the idea to scale down known radio wave technologies to the optical regime. One thoroughly investigated example of this approach are metallic nanoantennas which employ plasmonic resonances to couple localized emitters to selected far-field modes. While metals can be treated as perfect conductors in the microwave regime, their response becomes Drude-like at optical frequencies. Thus, plasmonic nanoantennas are inherently lossy. Moreover, their resonant nature requires precise control of the antenna geometry. A promising way to circumvent these problems is the use of broadband nanoantennas made from low-loss dielectric materials. Here, we report on highly directional emission from hybrid dielectric leaky-wave nanoantennas made of Hafnium dioxide nanostructures deposited on a glass substrate. Colloidal semiconductor quantum dots deposited in the nanoantenna feed gap serve as a local light source. The emission patterns of hybrid nanoantennas with different sizes are measured by Fourier imaging. We find for all antenna sizes a highly directional emission, underlining the broadband operation of our design.
	\end{onecolabstract}]   
		\begin{figure}[hhhh!tb]
		\centering
		\includegraphics[width=8cm]{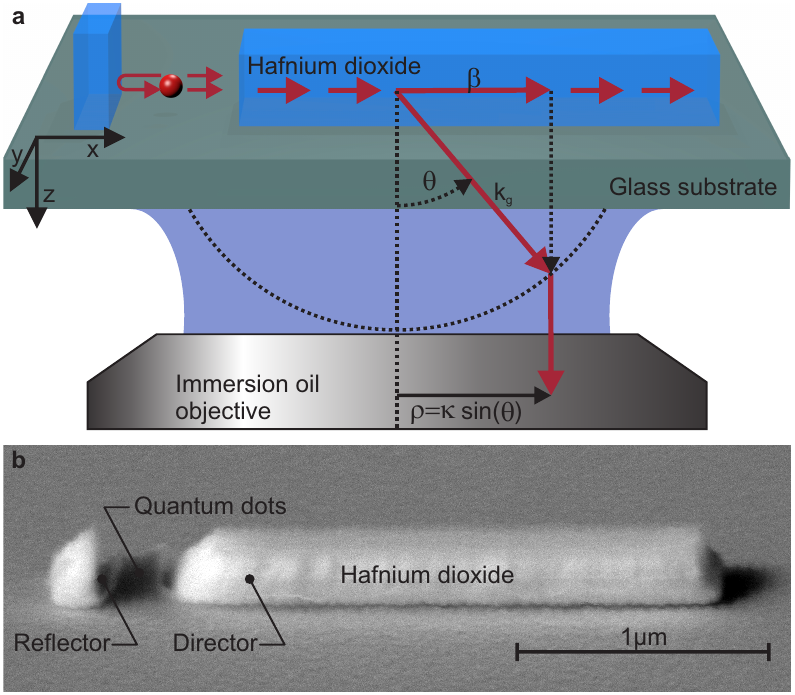}
		\caption{
			{\bf a} Schematic representation of the operating principle of the dielectric nanoantenna. The intensity distribution in the back-focal plane of the collecting objective is related to the angular distribution of emitted light by the sine-condition.
			{\bf b} Scanning electron micrograph of a Hafnium dioxide nanoantenna. Quantum dots (not visible) are deposited into the feed gap between director and reflector.}
		\label{fig:SEM_Hafnium}
	\end{figure}
	Nanoantennas have become valuable elements of the photonics toolbox to control and manipulate light on the nanoscale \cite{Bharadwaj2009, Novotny2011, Eisler2005}. They allow for an efficient interconversion of localized excitations and propagating electromagnetic waves \cite{Biagioni2012, taminiau2008optical}. In receiving mode, nanoantennas can locally increase the light intensity by several orders of magnitude \cite{lal2007nano, Maier2007, Krasnok2013}. This property can be used for the efficient excitation of quantum emitters \cite{pfeiffer2010enhancing, urena2012excitation} and to boost nonlinear effects \cite{Linnenbank2015, metzger2015strong, Schumacher2011,Wang2005, ghenuche2008spectroscopic}.
	In transmitting mode, coupling of quantum emitters to nanoantennas allows for the control of the emission properties\cite{kuhn2006enhancement,Curto2010, Hoang2015, dregely2014imaging,lee2011planar}. For instance, Curto {\it et al} reported on a highly directional plasmonic Yagi-Uda antenna\cite{Curto2010} and Lee \textit{et al.} demonstrated a planar dielectric antenna with near unity collection efficiency\cite{lee2011planar}.
	
	Like their microwave counterparts, nanoantennas can be categorized based on their functional principle into two large groups: (i) resonant antennas and (ii) nonresonant traveling wave antennas. So far, most research has focused on resonant nanoantennas based either on plasmonic resonances in metals \cite{Farahani2005, Hoang2015, Muskens2007, Knight2012} or on Mie resonances in high-refractive index dielectrics\cite{zhao2009mie, Staude2013,fu2013directional}. The latter offer the prospect of reducing dissipative losses while still providing  large resonant enhancements of the electromagnetic near field. A recent review on optically resonant dielectric nanoantennas can be found in reference \cite{Kuznetsovaag2472}. Moreover, dielectric antennas have been used in dielectric gradient metasurfaces as scattering elements \cite{Lin298}. In contrast to this, traveling wave antennas operating at optical frequencies have been studied considerably less.  However, there is a growing interest in transferring the traveling wave concept to higher operating frequencies in order to achieve non-resonant broadband operation \cite{Monticone2015,Shegai2011,Esquius-Morote2014}. 
	
	Leaky-wave antennas are a subset of traveling wave antennas that emit radiation over the whole length of a non-resonant guiding structure supporting the traveling wave \cite{Frezza2007}. In the case of a leaky-wave antenna with uniform cross-section, the phase velocity of the guided wave has to be larger than the velocity of light in the medium, into which the wave is radiated. The beam direction $\theta _\mathrm{beam}$ measured from the optical axis can be estimated (see Fig.~\ref{fig:SEM_Hafnium}a) by ${\sin \left( \theta _\mathrm{beam} \right)=\beta/k_\mathrm{g}}$, where $\beta$ is the propagation constant of the leaky mode for the given cross-section and $k_\mathrm{g}$ the wave number in the medium. 
	{The propagation constant $\beta$, and, hence, the beam direction can be controlled by designing the cross-section of the guiding structure.} The finite length of the wave guide as well as the radiation losses give rise to side lobes. The complete radiation pattern in the far field can be obtained by solving the Fraunhofer diffraction integral of the aperture distribution \cite{Frezza2007}. 
	
	In this letter, we report on a hybrid, dielectric, leaky-wave wave antenna for optical frequencies with high directivity. Our antenna design\cite{Hildebrandt2014} consists of only two simple dielectric building blocks deposited on a glass substrate. The total length of  the two dielectric building blocks is approximately three times the free space operation wavelength. The design can be easily adapted to various low-loss dielectric materials. Moreover, its non-resonant nature makes our antenna design inherently robust against fabrication imperfections and guarantees broad-band operation.
		\begin{figure}[ht]
		\centering
		\includegraphics[]{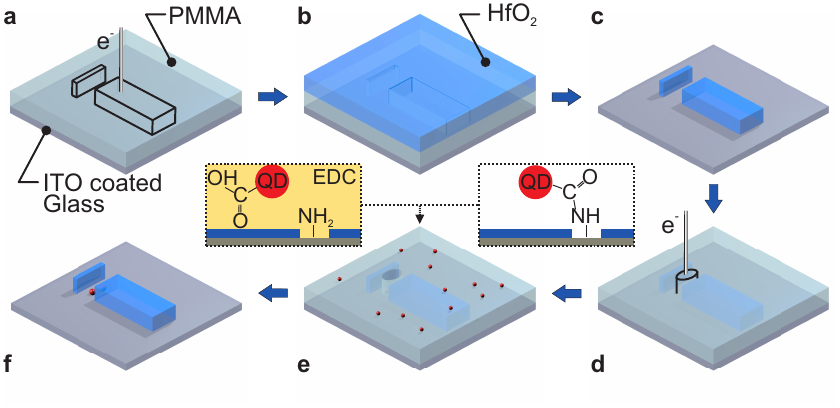}
		\caption{Fabrication of hybrid dielctric nanoantennas on ITO coated substrate: {\bf a} Geometry of dielctric nanoantenna is defined by electron beam exposure in PMMA. {\bf b} After development HfO$_2$ is evaporated. {\bf c} A lift-off process removes the PMMA and residual HfO$_2$ and produces the dielectric nanoantenna. {\bf d} In a new PMMA layer a patch in the feed gap of the antenna is exposed. {\bf e} After development an aqueous solution is applied to the template and a chemical link between QDs and ITO is moderated by EDC. {\bf f} A second lift-off removes the PMMA and residual QDs and the hybrid dielectric nanoantenna consists of the HfO$_2$ structure and QDs in the feed gap.
		}
		\label{fig:fabrikation}
	\end{figure}
	The leaky-wave antennas consist of Hafnium dioxide (HfO$_2$) nanostructures deposited on a microscope cover slip and use colloidal semiconductor quantum dots (CdSeTe) as fluorescent elements. HfO$_2$ has been chosen as the dielectric material for the antennas because it combines  a relatively large refractive index \cite{al2004characterization} (n=1.9) with very small absorption losses\cite{fadel1998study} at the emission wavelength of the quantum dots.
	
	The fabrication scheme is shown in Fig.~\ref{fig:fabrikation} and starts by performing electron beam lithography (EBL) on a standard microscope cover glass with refractive index $n_\mathrm{g}=1.52$:
	The sample substrate is a microscope cover glass coated with an 8-nm thick layer of Indium Tin oxide (ITO) to provide sufficient conductivity for the EBL process. As a lithography resist we use a double layer of Poly(methyl methacrylate) (PMMA) spin coated onto the substrate. The sublayers consist of 260 nm PMMA with 600 k molar mass and 200 nm PMMA with 950 k molar mass. The geometries of the antennas are written with a standard EBL system. A 180 nm thick film of HfO$_2$ is evaporated by electron-beam evaporation. During the deposition the temperature of the sample is kept under $\unit[100]{^\circ C}$.  The PMMA template and the residual HfO$_2$ is removed in a lift-off process, where the sample is submerged in $\unit[60]{^\circ C}$ warm N-Methyl-2-pyrrolidone (NMP) for $\unit[3.5]{h}$ .
	An electron micrograph of one of our dielectric antennas is shown in Fig.~\ref{fig:SEM_Hafnium}b. It consists of two $\unit[180]{nm}$ thick HfO$_2$ elements: The reflector has a footprint of  $\unit[180]{nm} \times \unit[785]{nm} $ and the director of $\unit[2200]{nm} \times \unit[600]{nm}$. They are separated by a  $\unit[260]{nm}$ wide feed gap.
	
	Colloidal semiconductor quantum dots (CdSeTe quantum dots, ZnS shell, Qdot 800 Carboxyl Quantum Dots, Thermo Fisher Scientific) with an emission wavelength of $\lambda_\mathrm{QD}=\unit[780]{nm}$ are used as the feed element of the hybrid dielectric nanoantenna. 
	The quantum dots (QDs) are coated with a polymer providing carboxyl surface groups and come in an pH buffered aqueous solution.
	They are precisely deposited from an aqueous solution into the feed gap of the antenna with the help of a second lithography step:
	The antenna sample is coated with a new PMMA layer and a second EBL step is applied.
	The $\unit[150]{nm} \times \unit[150]{nm}$ large area centered in the feed gap of each antenna, where QDs shall be deposited, is defined by exposure with the electron beam.
	After development, the PMMA film with the holes serves as a template for the subsequent surface functionalization. For this purpose, the sample is placed for an hour in a solution of 10 \% (3-Aminopropyl)triethoxysilane (APTES) in isopropyl alcohol to silanize the ITO layer in the unveiled patches. Next, 1-Ethyl-3-(3-dimethylaminopropyl) carbodiimide (EDC) is added to the QD solution and the substrate is immersed for two hours with constant stirring in this solution. EDC acts as an activating agent that mediates the link between the carboxyl surface groups of the QDs and the silanized substrate. After rinsing the substrate with deionised water, the PMMA mask is finally removed in a second lift-off process and the QDs stick to the modified surface areas in the feed gaps of the antennas. 
	
	The operating principle of the hybrid nanoantenna is shown in Fig.~\ref{fig:SEM_Hafnium}a and can be qualitatively understood as follows: The fluorescence of the quantum dots excites a leaky mode in the director by end-fire coupling. Light propagating along the director is continuously coupled to radiating modes in the substrate and emitted into the glass under an angle ${\sin \left( \theta _\mathrm{beam} \right)=\beta/k_\mathrm{g}}$ relative to the substrate normal, i.e., the optical axis. 
	Hence, when designing the antenna for a specific emission angle $\theta _\mathrm{beam}$, one has to consider both the geometry and the refractive index of the director as well as the refractive index of the substrate.
	Obviously, the condition $\beta < k_\mathrm{g}$ must be met since otherwise the director acts as a simple ridge waveguide and the leaky wave emission is prohibited.
	Emission into the air is prohibited since the phase velocity of the guided wave is smaller than that of light in air. To increase the gain of the antenna, the reflector redirects fluorescence emitted in the backward direction.
	This qualitative explanation can serve as a starting point for designing an antenna with a specific emission angle $\theta _\mathrm{beam}$.
	In the first step, the director cross section is chosen such that the leaky mode in the director has the appropriate propagation constant $\beta$. For this purpose, it is sufficient to perform a modal analysis with a 2D eigenvalue solver. 
	Next, 3D numerical calculations are used to iteratively improve the directivity of the antenna by varying the other geometry parameters, i.e., the length of the director, the size and position of the reflector.
	
	In the optical experiments, a blue pump laser ($\lambda=\unit[450]{nm})$ is focused by a high-numerical-aperture objective ($100 \times$ magnification, NA=1.49) through the substrate onto a single antenna to excite the quantum dots in the feed gap. 
	The fluorescence emitted by this hybrid antenna is collected with the same objective and separated from reflected pump light by a dichroic mirror and a series of optical filters.
	For our aplanatic objective lens, the spatial intensity distribution in the back-focal plane of the objective is related to the angular distribution of the collected light by the sine condition. A lens creates a real image of the back-focal plane on a scientific complementary metal-oxide-semiconductor (sCMOS) camera. 
	Thus, an emission angle $\theta$, measured with respect to the optical axis, corresponds to a distance $\rho=\sqrt{x_\mathrm{cam}^2+y_\mathrm{cam}^2}$ from the optical image center:
	\begin{equation}
	\rho=\kappa\sin(\theta) \quad,
	\end{equation}
	where $\kappa$ is the conversion factor.
	The biggest angle $\theta_\mathrm{NA}$ that can be still collected with the objective and hence be observed on the camera corresponds to the radius of a ring with ${\rho_\mathrm{NA}=\kappa \mathrm{NA}/n_\mathrm{g}}$.
	This relation is used with the known NA to determine the conversion factor $\kappa$.
	
	Antenna data is commonly represented in spherical coordinates ${(\theta,~\varphi)}$, where $\theta$ is the polar and $\varphi$ the azimuthal angle. 
	In our analysis, we choose the orientation such that the antenna axis points in the ${(\theta=90^\circ,~\varphi=0^\circ)}$ direction and the optical axis corresponds to the ${(\theta=0^\circ)}$ direction. The data is presented in this letter with a linear $\theta$ axis and not in the pseudo momentum space of the camera chip.
	So, the transformation from the $x_\mathrm{cam}$ and $y_\mathrm{cam}$ coordinates of the camera chip to spherical coordinates reads:
	\begin{eqnarray}
	\theta&=&\arcsin \left(\frac{\sqrt{x_\mathrm{cam}^2+y_\mathrm{cam}^2}}{\kappa} \right),\\
	\varphi &=& \arctan \left( \frac{y_\mathrm{cam}}{x_\mathrm{cam}}\right) \quad .
	\end{eqnarray}
	To investigate the polarization of the antenna signal, we place a linear polarizer as an analyzer in front of the camera.
		\begin{figure*}[ht]
		\centering
		\includegraphics[width=13.8cm]{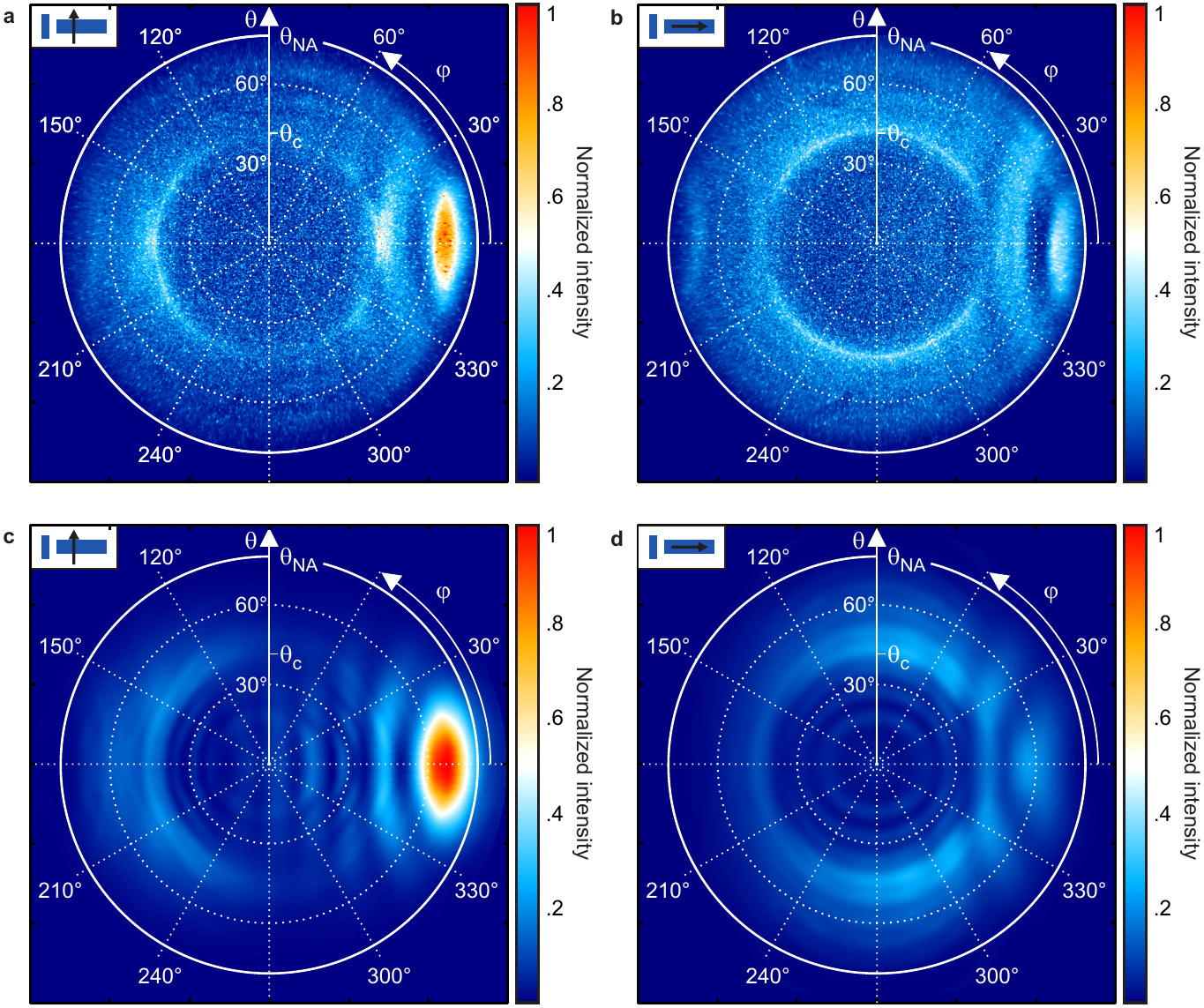}
		\caption{{\bf a} and {\bf b} Measured angular intensity distributions of an hybrid dielectric nanoantenna for two different analyzer settings (see inset) normalized to the same value. The main lobe maxima in both distributions are at $\theta_\mathrm{max}=70^\circ$. The central ring-like feature is attributed to dipoles not coupled to the antenna emitting directly into the substrate. The white circles at $\theta_{NA}=79^\circ$ mark the experimentally accessible angular range.
			The calculated intensity distributions for the corresponding analyzer settings in {\bf c} and {\bf d}, show a similar behavior as the experimental data.
		}
		\label{fig:MEssung_Theorie_RL}
	\end{figure*}
	Figure~\ref{fig:MEssung_Theorie_RL}a depicts the normalized angular intensity distribution emitted by the hybrid dielectric nanoantenna shown above. Here, the analyzer axis is set perpendicular to the antenna axis, i.e., we record the emission of a TE-polarized leaky mode (see inset). 
	The hybrid antenna shows a highly directional emission with a strong main lobe at ${(\theta_\mathrm{max}=70^\circ,~\varphi_\mathrm{max}=0^\circ)}$. 
	This lobe has a full width at half maximum of ${\Delta \theta_\mathrm{max}=(9 \pm 2 )^\circ}$ and ${\Delta \varphi_\mathrm{max}=(24 \pm 4 )^\circ}$. About 6~\% of the total collected intensity is confined in the main lobe. Additional concentric side lobes around the main lobes are visible.  
	A reference measurement (not shown) with a bare quantum dot sample indicates that the weak circular feature at $\theta\approx\theta_\mathrm{c}=41.1^\circ$ can be attributed to uncoupled quantum dots, which preferentially emit at the critical angle between air and glass \cite{novotny2012principles}.
	
	The directivity $D$ of an antenna is defined as the ratio of the peak intensity and the intensity averaged over all directions as observed in the far field \cite{Kraus2002}.
	The collection angle in our experiment is limited by the NA of the microscope objective, i.e., light emitted by an angle larger than $\theta_\mathrm{NA}=79^\circ$ is not detected. 
	As a result, a part of the intensity distribution is cut off. 
	With this restriction in mind, the directivity of the antenna over the measured part of the distribution can be estimated to be $D=\unit[12.5]{dB}$. 
	We additionally use the front-to-back ratio (F/B), defined \cite{Curto2010} as the intensity ratio between the maximum at (${\theta_\mathrm{max},~ \varphi_\mathrm{max}}$) and the opposing point (${\theta_\mathrm{max},~\varphi_\mathrm{max}+180^\circ}$), to quantify the directional performance of our hybrid antenna.
	The F/B value of the dielectric nanoantenna measured here is $\unit[12]{dB}$.
	
	The angular intensity distribution for the analyzer axis parallel to the antenna is shown in Fig.~\ref{fig:MEssung_Theorie_RL}b. It is normalized to the same value as the data discussed above. The peak intensity as well as the directivity ($D=\unit[9]{dB} $) are in this case smaller than that recorded for the perpendicular analyzer setting (compare Fig.~\ref{fig:MEssung_Theorie_RL}a and b). 
	A plausible explanation for these observations is that the coupling of the quantum dots to the TM-polarized leaky  mode is less efficient. This interpretation is consistent with numerical calculations.  
	
	To support our experimental findings, numerical calculations based on finite integration technique (FIT) using CST Microwave Studio were performed \cite{weiland1977discretization}.
	A single dipole in the feed gap served as the fluorescent element and three different perpendicular dipole orientations along the coordinate axes (see Fig.\ref{fig:SEM_Hafnium}a) were assumed in successive calculations.
	For each dipole orientation, the far-field intensities for the two analyzer settings used in the experiments were evaluated separately.
	Finally, the intensities of the three dipole-orientations are summed for each analyzer setting.
	With this procedure, we simulate the ensemble of QDs with random dipole orientations as used in the experiment. 
	The calculated intensity distributions for both analyzer settings are shown in Fig.~\ref{fig:MEssung_Theorie_RL}c and \ref{fig:MEssung_Theorie_RL}d.
	They feature the same main and side lobes as the experimental data.
	The corresponding directivities for the analyzer in the $y$- and $x$-direction are $D=\unit[14.05]{dB}$ and  in $D=\unit[8.347]{dB}$, respectively.
	A detailed analysis of the different dipole orientations shows that there are two main contributions to the main lobes: (i) the dipole oriented along the $y$-direction couples primarily to the TE leaky mode and (ii) the dipole oriented along the $z$-direction predominately excites the TM leaky mode.
	A comparison of these two cases shows that the coupling efficiency to the TM mode is smaller, resulting in a lower directivity for light polarized along the antenna axis. 
	
	To assess the overall performance of the dielectric antenna, we have performed a more detailed analysis of the calculated angular intensity distribution. The overall emitted intensity, $I_\mathrm{total}$, is obtained by integrated the far field intensity over the full  $4\pi$ solid angle. The intensity collected by the microscope objective, $I_\mathrm{NA}$, is calculated by integrating the intensity within the collection angle of the objective, i.e., $\theta\le\theta_\mathrm{NA}$. For the dipole oriented along the $x$-, the $y$- and the $z$-direction, we find that the ratio $I_\mathrm{NA}/I_\mathrm{total}$ is 73~\%, 80~\% and 91~\%, respectively. These values suggest that a large fraction of the overall intensity is indeed observed in the experiment. A corresponding analysis shows that the main lobe contains 12~\% of the intensity collected by the microscope objective, $I_\mathrm{NA}$ (averaged over the three dipole orientations). This value is approximately twice as large as the measured value. We attribute this to the fact that in the experiment not all QDs couple ideally to the antenna mode due to a displacement from the optimized position assumed in the calculations. Moreover, the electron micrograph of the antenna reveals (see Fig.~\ref{fig:SEM_Hafnium}b) that the sidewalls of the director are not perfectly smooth. Hence, scattering from this surface roughness constitutes another loss channel for our antennas.
	
	From the calculated field distributions, we determined\cite{JohnsonFDTD} the Purcell factor to be $F_{\textrm P}=1.02$. 
	A Purcell factor of order unity is not unexpected for the considered antenna design. First, the antenna is broadband (see below), i.e., it has a small quality factor. Secondly, the electromagnetic near field is not confined to a deep sub-wavelength volume. Moreover, the directional emission requires destructive interference for one-half space which tends to reduce the Purcell factor\cite{Koenderink2017}.
		\begin{figure}[h!tb]
		\centering
		\includegraphics[width=8cm]{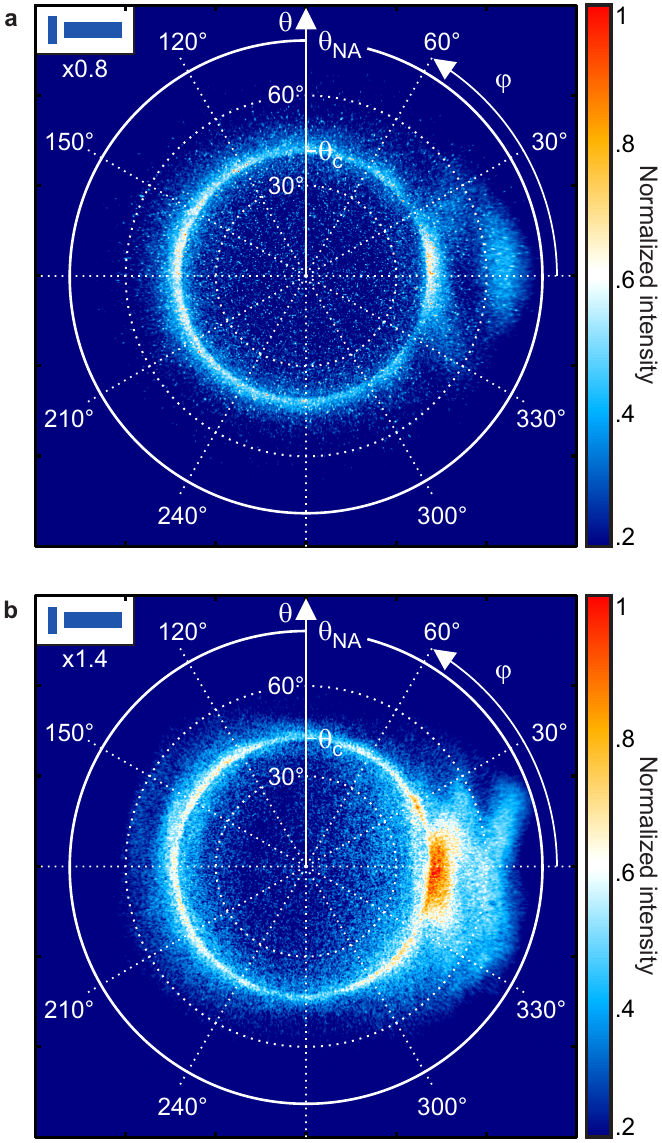}
		\caption{Angular intensity distribution of dielectric antennas without analyzer with footprints (\textbf{a}) $0.8 \times$  and (\textbf{b}) $1.4 \times$ as large as the original design. The white circles mark the experimentally accessible angular range.}
		\label{fig:messung_VS_VB}
	\end{figure}
	A main advantage of non-resonant antennas is their high bandwidth and robustness against fabrication imperfections. 
	Hence, we anticipate that the variation of the antenna dimensions will result in a different beam direction but not a total loss of the antenna's functionality.
	We measured the angular intensity distribution of antennas whose lateral dimensions (footprints of the director and the reflector as well as the gap size) were scaled by a factor $0.8 \times$   and $1.4 \times$, respectively, relative to the original design. The height was not scaled. As anticipated, both antennas still show directional emission (see Fig. \ref{fig:messung_VS_VB}a and b) . 
	This behavior agrees well with the numerical calculations, which predict a plateau of high directivities for a broad range of the widths and lengths around the original design \cite{Hildebrandt2014}. 
	For instance, a reduction of the gap size by $200$~nm decreases the directivity by only 2 dB. Moreover, the directivity does not critically depend on the exact position of the dipole. Numerical calculations show that the directivity stays above 10 dB if the dipole is located within a 150 nm $\times$ 150 nm large rectangle which includes the optimal dipole position. This rectangle corresponds to the patch defined in the second lithography step in our sample fabrication process.
		\begin{figure}[ht]
		\centering
		\includegraphics[width=8cm]{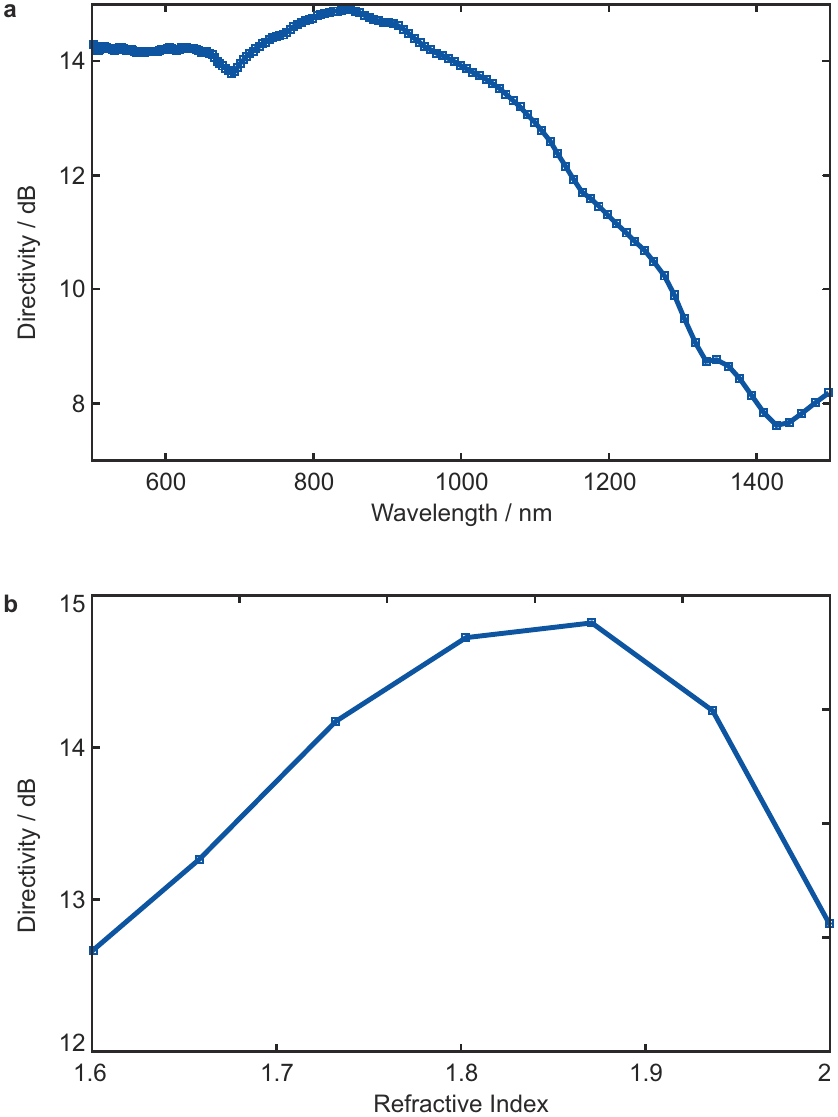}
		\caption{ (\textbf{a}) Calculated directivity of the dielectric nanoantenna as a function of the emission wavelength. 
			(\textbf{b}) Calculated directivity of the dielectric nanoantenna as a function of the refractive index of the reflector and director. Blue lines added to guide the eye.}
		\label{fig:Directivity_Theory}
	\end{figure}
	To further substantiate the claim of broadband operation, additional numerical calculations where performed, in which we varied the operation frequency of the exciting dipole. All other parameters were kept fixed. The resulting directivity as a function of excitation wavelength is shown in Fig.~\ref{fig:Directivity_Theory}a. 
	These calculations clearly support our claim: The directivity is larger than $\unit[10]{dB}$ in the range from $\unit[500]{nm}$ to $\unit[1200]{nm}$ wavelength.
	
	Our design is not only robust against deviations of the fabricated geometry from the specifications but also tolerates variations of the refractive index of the reflector and the director without reoptimization of the geometry. Figure~\ref{fig:Directivity_Theory}b depicts the calculated directivity vs refractive index of the two elements for a $y$-oriented dipole. The geometry parameters have not been changed. The directivity takes its maximum value for the refractive index $n=1.9$, i.e, the refractive index for which the antenna has been designed. Notably, the directivity stays above 10 dB even for considerable variations of $n$ without accompanying optimization of the geometry.
	
	It is instructive to compare the performance of our antenna with prominent previous work. 
	As stated above, the forward to backward ratio F/B of our dielectric nanoantenna is $\unit[12]{dB}$. This value is quite competitive in comparison with highly directive plasmonic nanoantennas. For instance, in the seminal work on plasmonic Yagi-Uda antennas a F/B value of $\unit[6]{dB}$ has been reported\cite{Curto2010}.
	Another important parameter is the photon collection efficiency, which is defined as the fraction of the total emitted power that is captured by the used far-field collection optics.
	Lee \textit{et al.} \cite{lee2011planar} reported on a 96 \% collection efficiency for a planar dielectric antenna and a NA=1.65 microscope objective. From the numerical calculations, we determine a collection efficiency of 81~\% for our system (dielectric antenna and NA=1.49 microscope objective).
	When comparing these values one should keep in mind that the two antennas give rise to quite different angular emission patterns. The emission of the planar dielectric antenna is evenly distributed over a ring centered around the optical axis while our hybrid dielectric antenna features a single pronounced main lobe. Moreover, our antenna selectively couples to emitters in the feed gap while in the case of the planar dielectric antenna the lateral position of the emitter is non-critical. Whether this is an advantage or a disadvantage depends on the respective experiment.
	Another promising approach to achieve efficient broadband emission into a free-space beam is to directly embed the emitters either into a microcavity \cite{somaschi2016near} or into a high dielectric taper structure \cite{munsch2013dielectric}. Unfortunately, this approach is not compatible with all types of quantum emitters, e.g., it is not straightforward to incorporate dye molecules or nanocrystals in these epitaxially grown structures. For such emitters, our dielectric antenna might be a promising alternative. 
	
	In conclusion, we have fabricated and characterized hybrid dielectric nanoantennas for the optical regime.
	The antennas exhibit highly directional emission.
	Experiments with different antenna sizes  indicate the broadband operation of our nanoantenna design.
	These characteristics make the hybrid antenna a promising candidate for future applications. We envision that the dielectric antenna in combination with a single quantum emitter may be used as a highly directional single-photon source without inherent losses. By placing the dielectric antenna into a liquid crystal cell, the beam direction can potentially be tuned electrically.
	
	The authors declare no competing financial interest.
	
	\section*{Acknowledgements}
		S.L. and M.P. acknowledge financial support through DFG TRR 185 and by the German Federal Ministry of Education and Research through the funding program Photonics Research Germany (project 13N14150). A.H., J.F., and T.Z.  acknowledge financial support through DFG TRR 142 and DFG GRK 1464.

	\bibliography{bib_diel}

	%\begin{figure}[ht]
	%	\centering
	%	\includegraphics[]{FG_NL_Title_Pic.pdf}\\	
	%    TOC Graphic
	%	%\caption{This image is supposed to accompany the abstract.}
	%	\label{fig:AbstractArt}
	%\end{figure}

\end{document}